# The effect of micro-changes in the pluck trajectory on the sound of an acoustic guitar

**Marek PLUTA**, **Jan JASIŃSKI**, **Daniel TOKARCZYK**, **Julia GRYGIEL**

AGH University of Krakow, al. Mickiewicza 30, Krakow, Poland

**Corresponding author:** Marek PLUTA, email: pluta@agh.edu.pl

**Abstract** This study explores how micro-changes in the plucking trajectory of a guitar pick influence the sound of an acoustic guitar. Using a state-of-the-art robotic plucker, a series of measurements has been performed, where the plectrum was moved towards the instrument by a step of 192 micrometers, resulting in an increased attack depth. It has been analysed how the effect of these changes is reflected in loudness, timbre, harmonic content and how the sound progresses during decay. This methodology has been repeated for guitar plectra made from six different materials to investigate how the pick itself influences the effect of a change in the plucking trajectory. The results of the study show that at a low depth the string is not fully excited resulting in weak and markedly altered sound. The range of this effect changes with the mechanical properties of the plectrum material. After this range an increase in depth results in an increase in sound loudness, a decrease in inharmonicity and noisiness and a shift in timbre where the sound becomes fuller in low frequencies and rougher. Presented findings help to understand the nuanced relationship between plucking trajectory and acoustic output. They provide important insights regarding the importance of plucking in guitar testing methodologies, showing that the mechanics of plucking must be taken into account when conducting and interpreting results of guitar testing.

## 1. Introduction

One of the reasons for studying musical instruments is to evaluate an impact of modifications in instrument's geometry, material, or playing technique on the properties of its sound. This kind of study involves producing a sound with an instrument in question, and recording it either with microphones, or other transducers that allow to study vibrations of chosen elements. The key problem is choosing a proper mechanism for producing sound. Depending on studied phenomenon a compromise needs to be achieved between stability and repeatability of the mechanism and its flexibility in altering playing parameters.

In chordophones a string can be excited in one of three ways: it can be plucked, struck, or bowed [1]. This classification is applied in numerical models, but in the case of real instruments the division might be less clear, particularly in the case of plucking, which often has some properties of striking due to a non-zero velocity of the plucking element. Moreover, material and geometric properties of various plucking elements as well as uncertainties related to mounting an instrument on a test stand can limit precision of establishing an actual plucking position and its spatial distribution, as well as the distance at which the string is released, both of which directly affect properties of the produced sound.

Even in the case of a guitar, which is one of the most popular, and extensively studied instruments, the above mentioned issues remain open. Despite applying various mechanisms to pluck a guitar string in numerous studies [2–6], the sensitivity of an instruments' sound to geometric properties of the plucking process has not been evaluated in adequate detail. It leaves a question regarding the precision that should be required from such mechanisms in order to reliably study subtle alterations to the sound of an instrument caused, for example, by variations in playing techniques, or by replacing some of its elements with counterparts made of different materials.

The study presented in this manuscript attempts to provide a new and precise data on the effect of sub-millimeter changes in the pluck trajectory on selected properties of sound of an acoustic guitar. A state-of-the-art robotic plucker [7] has been applied to produce series of plucks where the plectrum was shifted towards the instrument by a step of 192 micrometers, which resulted in an increased attack depth, displacing a string farther before being released. Produced sounds have been recorded and analysed to learn how the effect of these changes affects loudness, timbre, harmonic content and how the sound evolves during a decay phase. In order to investigate how the pick itself influences the change caused by plucking trajectory, the study has been carried out for guitar plectra made from six different materials. Presented findings help to understand the nuanced relationship between plucking trajectory and acoustic output. They provide significant insights regarding the importance of precision required from plucking mechanisms in guitar testing methodologies.

## 2. Impact of plucking mechanics on a timbre of sound

Plucking is the most basic way to excite a string. It is relatively straightforward to simulate in numerical models of instruments, and due to a fact that it is a default performance technique in the majority of string instruments, various aspects of plucking have been studied and discussed in literature.

Traube and Smith [8] developed a signal processing technique that allows to extract the plucking point on a string from a sound recording of a guitar, proving that the impact of plucking position on sound alone is large enough to acquire unambiguous information regarding useful details of a performance technique. Determining the plucking point allowed to formulate a method of detecting the fingering point. Together, they can be applied to produce transcriptions of a guitar music with performance details included. A different approach has been applied by Perez-Carrillo [9], who studied plucking using motion capture technique with high-speed cameras, solving the problem of visual occlusion of fingertip markers by applying rigid-body and flexible-body models to track the motion of strings and fingers. Performed analyses, including plucking position, timing, velocity, and direction, allowed to observe and discuss finger-string interactions.

Another factor that has an impact on the effect of a pluck is the thickness of a plectrum. The problem has been studied by Carral and Paset [10], who plucked a string of an acoustic guitar with three different plectra mounted on a dedicated plucking mechanism. Plucking trajectory remained the same, and the only variable was the thickness of the plectrum (0.46, 0.96 and 2.0 mm). Recorded sounds were analysed to obtain data regarding evolution in time of signal level, pitch, and spectral attributes. A similar study for an electric guitar has been carried out by Agrawal [11]. In this case three plectra (0.5, 1.0, and 1.5 mm) produced sounds that were analysed with regards to the signal level only. Here again, the only variable was the thickness of a plectrum. Plucking trajectory was not altered during the experiment. Altering a trajectory would require a more flexible and complex mechanism than these applied in both studies.

A test rig that allowed to vary a vertical plucking angle has been designed by Tahvanainen and Hochmuth [12], who applied it to study the effect of plucking angle on sound produced by a string of the Finnish kantele. Three plucking positions and four angles have been analysed. Obtained results show that vibrations of a kantele string have two vertical polarisations and a longitudinal vibration. Changes in angle affected the beating phenomenon occurring for string partials due to inharmonicity. A broader set of factors affecting timbre of a guitar sound with regards to certain genres of music has been studied by Ramsey and Moore [13]. Introduced alterations included the presence of a capo, various strings, plucking with fingers and plectra, as well as plucking position. The study aimed to improve musical expression capabilities as a main implementation of obtained results.

This brief survey shows various approaches to studying plucking mechanics and its impact on sound. In the majority of experiments the set of studied cases is relatively narrow due to limited capabilities of plucking mechanism. The same applies to studied variables that are changed in coarse steps. A more flexible and precise mechanism could reveal more details regarding studied phenomena. It would also allow for the study interdependence of different factors independently.

## 3. Experiment

The aim of the experiment was to observe the effect of sub-millimeter changes introduced to pluck trajectory on the sound of an acoustic guitar. In a series of plucks the distance between a tip of the plectrum and a top plate of a guitar was changed in 192 micrometer steps. Increasing plectrum attack depth resulted in displacing a string farther before being released. This effect would depend on plectrum thickness and material, therefore the measurements were repeated for different plectra. As a consequence two factors varied during the experiment: plectrum trajectory, and plectrum material.

### 3.1. Test stand

Measurements involving sub-millimeter adjustments required adequately precise mechanism to repeatedly position the plectrum, shift it to a desired depth, and pluck the string. The experiment has been carried out using a test stand equipped with a Cartesian coordinate robot and a set of studio condenser microphones.

The stand was installed in a large anechoic chamber in the Department of Mechanics and Vibroacoustics of the AGH University of Krakow.

A Cartesian coordinate robot used in the experiment is a device designed particularly for studying guitars (Fig. 1) [14]. It has three linear axes of control and can precisely move a string plucking mechanism inside a rectangular area of 500 mm along instrument strings, 250 mm along a top plate, perpendicular to strings, and 250 mm perpendicular to a top plate. The frame of the robot is made of V-slot aluminium profiles and uses dedicated carts. The drive is transferred by trapezoidal screws from NEMA 17 stepper motors controlled with DRV8825 drivers. Electronics assembly parts and the end effector with a mounting for guitar picks are printed from PLA. The minimal displacement along a single axis is 0.04 mm. Theoretical maximal velocity is 20 mm/s.

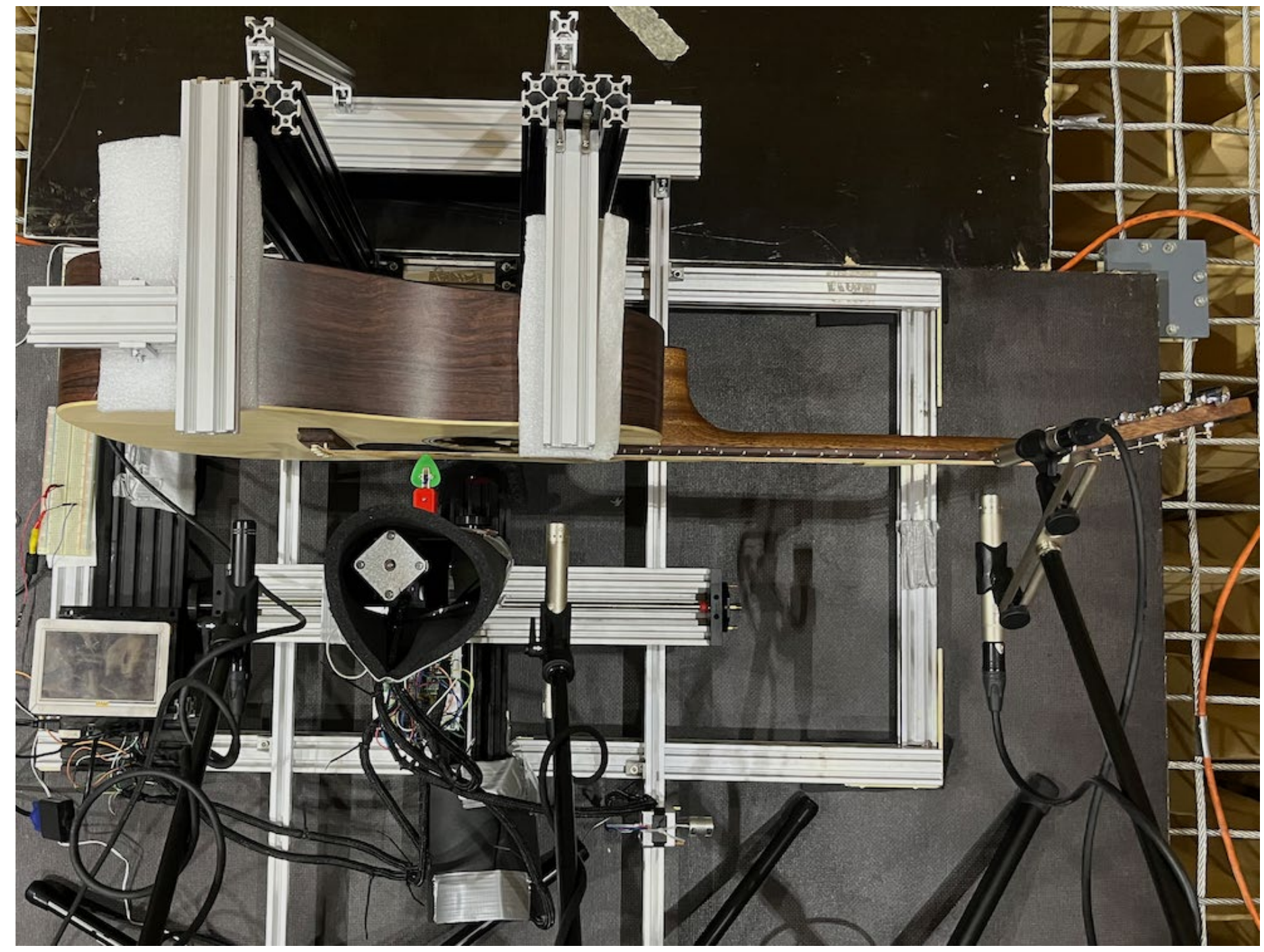

**Figure 1.** Test stand with Cartesian coordinate robot, Martin D-X2e acoustic guitar, four microphones, all set in anechoic chamber.

The sound produced by plucked string has been recorded with one RODE MP5 and three Studio Projects C4 microphones connected to Zoom F8n Pro recorder. MP5 was located 120 mm above the top plate of the instrument and pointed at the lower part of the body near the bridge. One C4 was pointed at the last fret, 120 mm above it, and two others, perpendicular to each other, were pointed at the second fret, 35 mm from the string axis, as shown in Fig. 1.

The plucking trajectory involves movements along two axes: horizontal $x$-axis perpendicular to a guitar top plate, and vertical $z$-axis parallel to a top plate and perpendicular to a string. Movement along $z$-axis is used to pluck a string and later to return to initial position. Movement along $x$-axis changes the distance between a plectrum and a top plate, and also allows to avoid plucking a string on a way back to plucking position. The string is always plucked from the same direction, with the same side of a plectrum. The diagram of subsequent stages of movement is presented in Fig. 2. A string-plectrum contact is achieved half the way along $z$-axis. After plucking, the robot stops for 30 s before moving again along $x$-axis, to allow acquiring a clear recording of vibrating string.

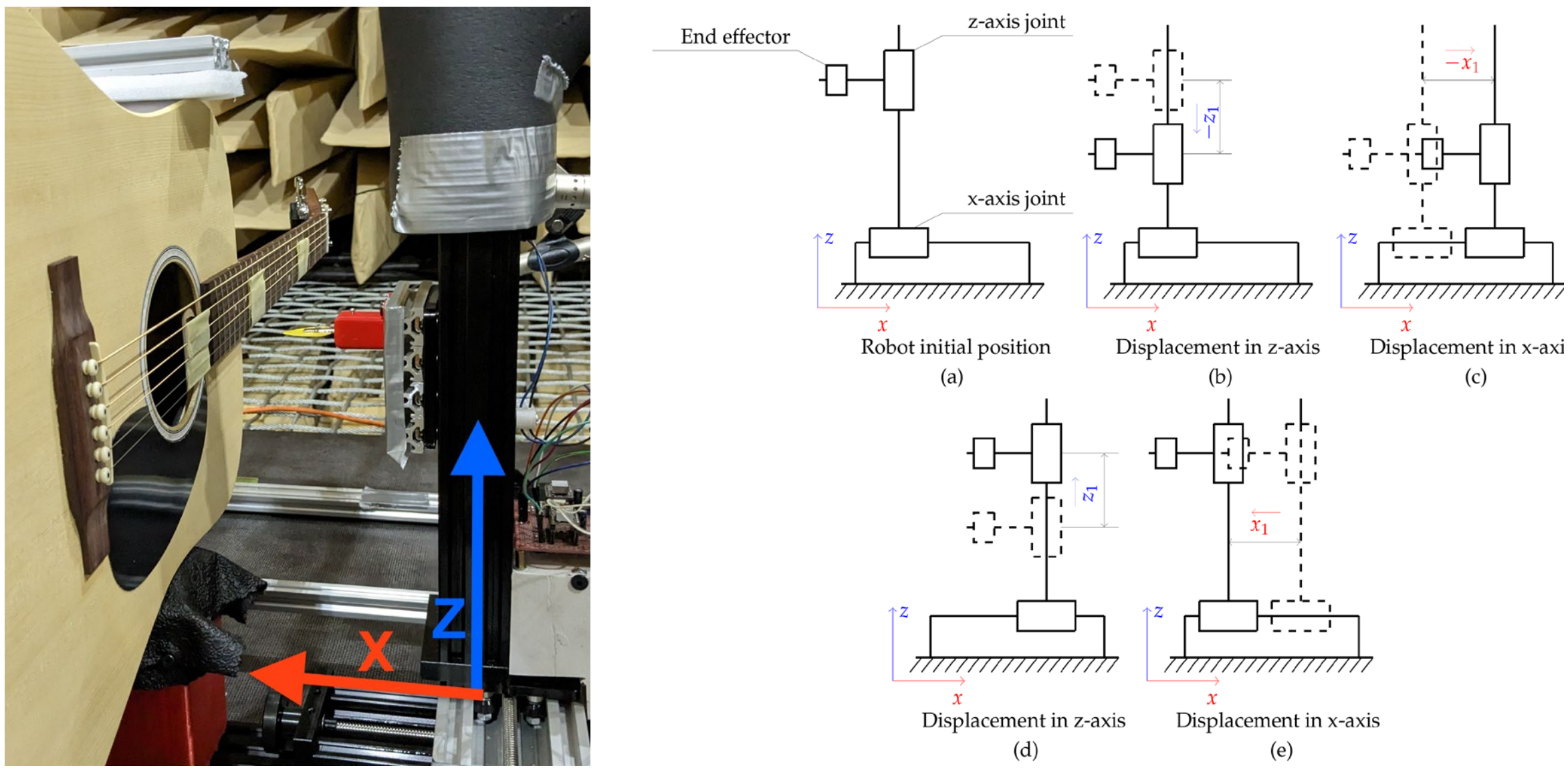

**Figure 2.** Relative positions of plectrum (yellow element with red mounting) and string on the left, and simplified diagram of the movement on the right.

### 3.2. Test procedure

In the experiment a total of eight plectra have been examined. Each plectrum was plucking a string with six trajectories differing in plucking depth, understood as a distance between a tip of a plectrum and the string edge closer to the robot, as shown in Fig. 3. In the first plucking position (I) the tip of a plectrum was barely touching a string. In the last one (VI) the tip still did not reach the opposite string edge. The distance between subsequent depths was 192 μm, with a string (E6) diameter 1.42 mm.

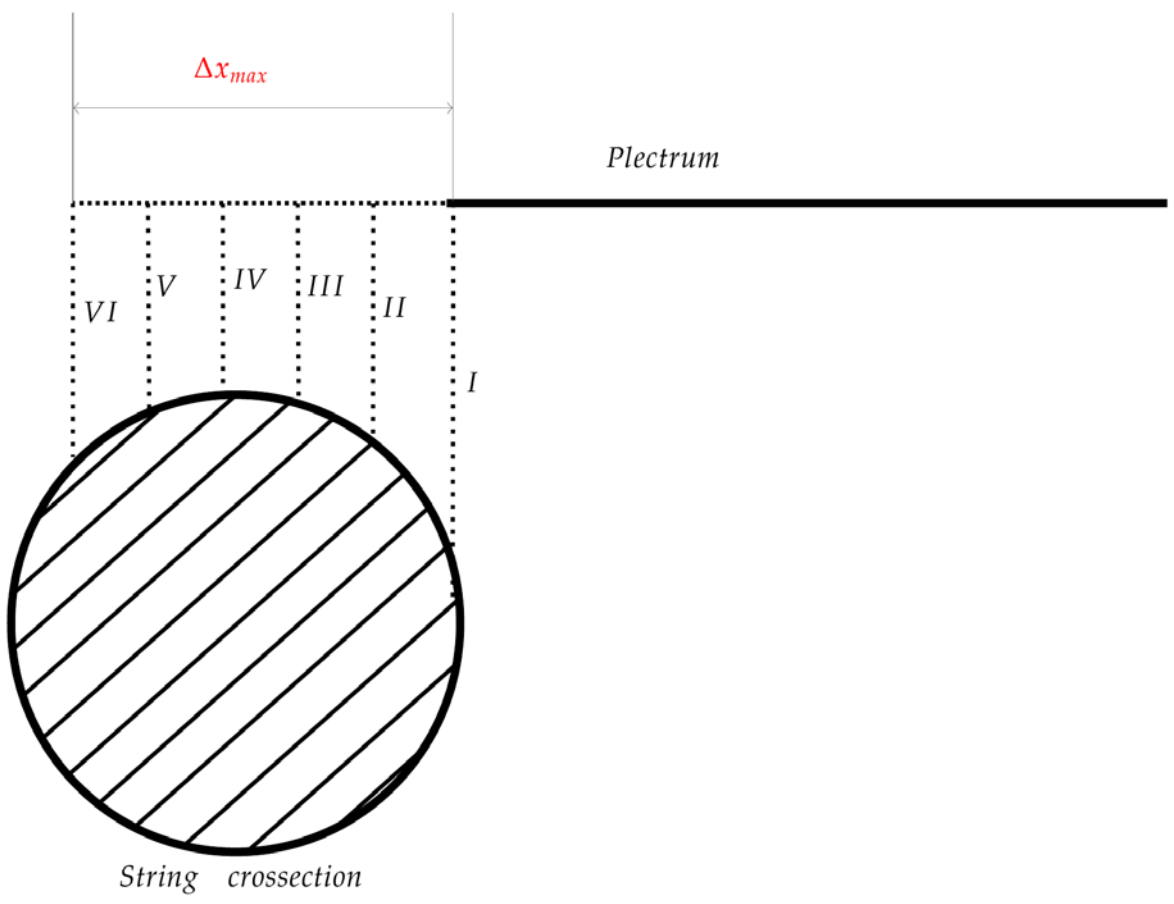

**Figure 3.** Six plucking depths (I–VI) presented in relation to the diameter of string E6 used in the experiment (1.42 mm); distance between subsequent positions was set to 192 μm, and a total distance $\Delta x_{\max}$ (between positions I and VI) was 960 μm.

For each combination of plectrum and plucking depth the robot performed 10 plucks, later to be averaged. Therefore a total of 480 plucks have been recorded (8 plectra, 6 depths, 10 repetitions each) for further analysis. The procedure consisted of the steps enumerated below:

1. Mount a plectrum.
2. Calibrate the robot.
3. Set an initial plucking depth.
4. Perform 10 plucks.
5. If (plucking position < VI) then increase depth and return to step 4.
6. Select next plectrum and return to step 1.

Additionally, for one selected plectrum steps 1–5 were repeated three times. The same plectrum was removed and remounted, and all the measurements were carried out again, in order to test repeatability of mounting conditions.

The calibration process (step 2 of the procedure) was performed to find the initial plucking position (I). The plectrum was moved to the assumed initial position along all axes. The subsequent steps of a stepper motor back were performed until losing contact between the plectrum and the string. One step of a stepper motor forward was performed to achieve the initial position with the plectrum barely touching the string.

### 3.3. Plectra

When a plectrum displaces a string, the moment of releasing it may depend on plectrum elastic and frictional properties. Therefore the same plucking depth can result in different initial amplitude and different modes of vibration present. Table 1 presents the properties of the guitar plectra used in the study (Fig. 4). The information was obtained from the official websites of the respective manufacturers [15, 16]. For felt and polycarbonate plectra, thickness data was not available, therefore, their thickness was measured manually using a calliper.

**Table 1.** Properties of the guitar picks used in the study.

| | Brand | Material | Thickness |
|---|---|---|---|
| 1. | Dunlop | Nylon | 0.67 mm |
| 2. | Dunlop | Nylon | 0.80 mm |
| 3. | Dunlop | Nylon | 0.94 mm |
| 4. | Dunlop | Nylon | 1.14 mm |
| 5. | Dunlop | Polycarbonate | 1.30 mm |
| 6. | Dunlop | Steel | 0.38 mm |
| 7. | Wedgie | Rubber | 3.1 mm |
| 8. | Dunlop | Felt | 3.2 mm |

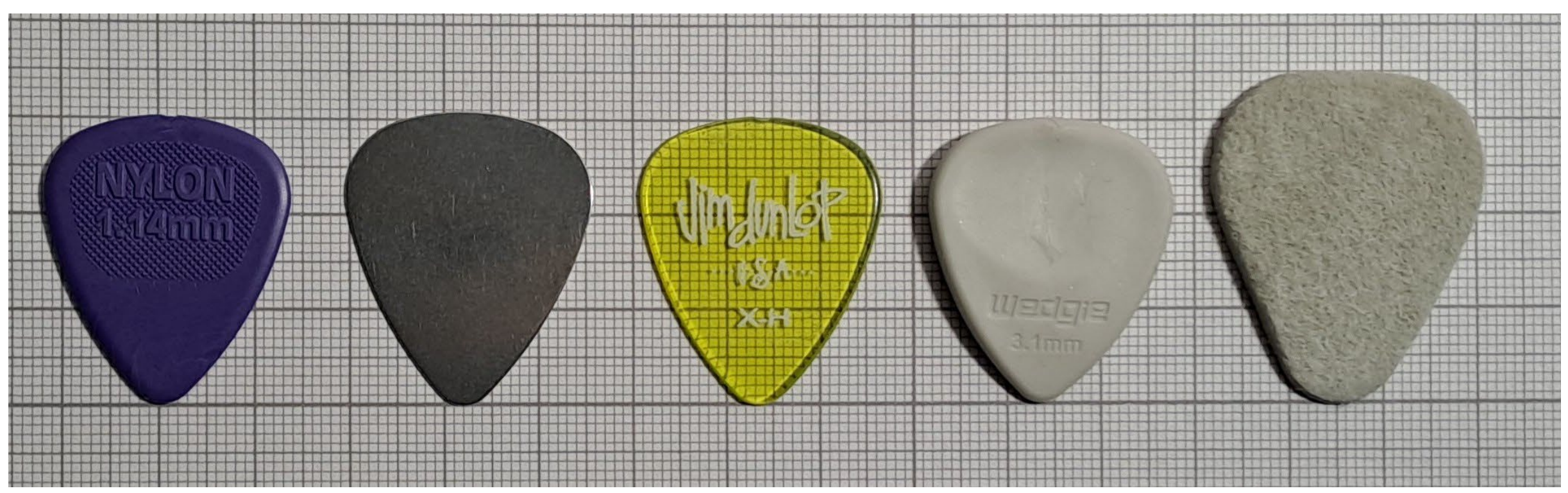

**Figure 4.** Comparison of pick sizes and shapes on millimeter graph paper.
From left to right: nylon, steel, polycarbonate, rubber and felt.

## 4. Results

The measurement recordings were parsed and for each pluck a set of signal representations and features were extracted. In this paper only a subset of results will be presented but the presented conclusions are based on a full analysis.

### 4.1. Spectrum analysis

To understand the changes caused by an increase in plucking depth it is worth observing the spectra of the recorded signals. Fig. 5 presents spectra recorded for each plucking position for nylon (0.8 mm thickness) and steel.

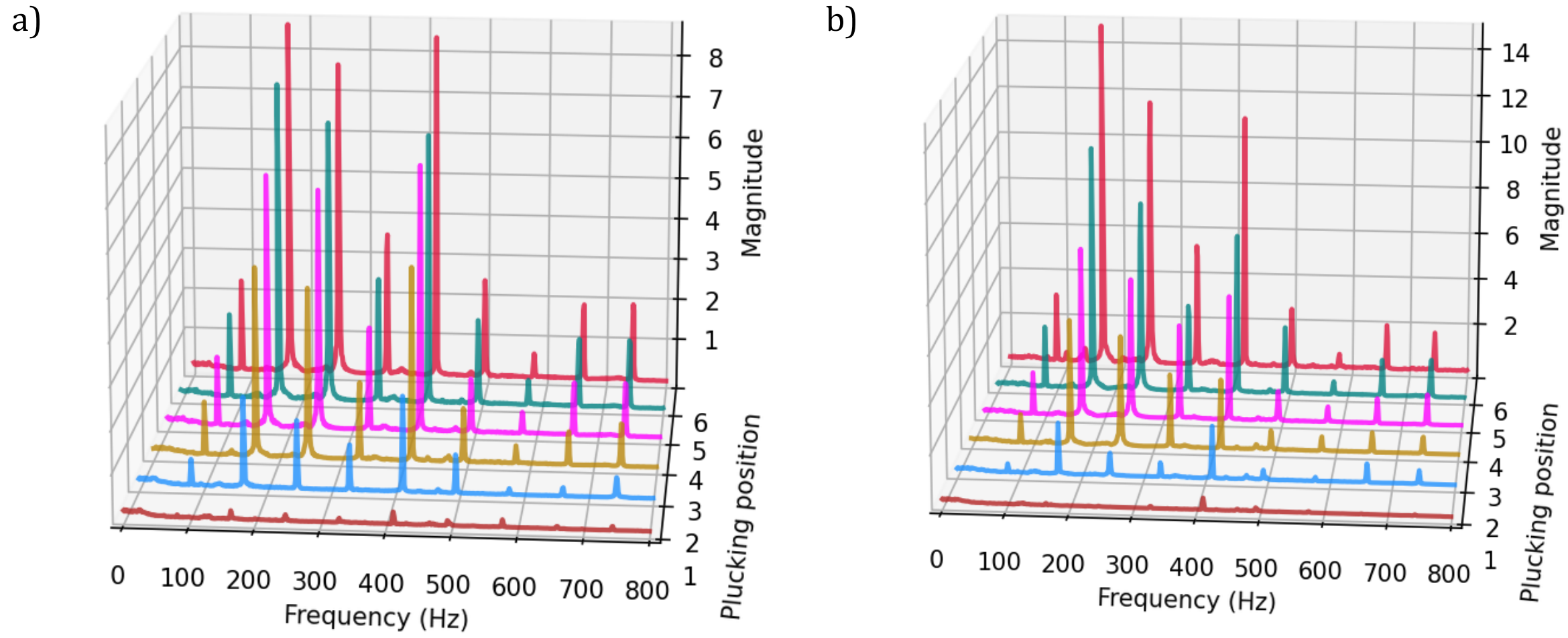

**Figure 5.** Comparison of average spectra for measurement series conducted for each plucking position with a a) nylon 0.8mm, b) steel plectrum.

The graphs show that for all materials an increase in depth understandably increases the magnitude of harmonics, but this rise is proportionally larger for low frequencies. This can be seen on Fig. 5b where in the first position frequencies below the fifth harmonic are not visible. Moving to the second position all harmonics become clear, but the fifth harmonic is still dominant. For all other depths the second harmonic rises and becomes the most prominent. This shows that an insufficiently light pluck fails to excite the lower harmonics of the string. It is also important to highlight that for both nylon and steel, starting from the third position and beyond, the ratios between the first nine harmonics remain mostly consistent, with the only significant change being a proportional increase in their magnitude.

### 4.2. Loudness and energy parameters

The analysis of an instrument's sound is commonly divided into the categories of loudness, duration and timbre [17], thus the first aspect worth investigating is the loudness and energy of the recorded signals. Fig. 6 shows the average loudness calculated in accordance with [18].

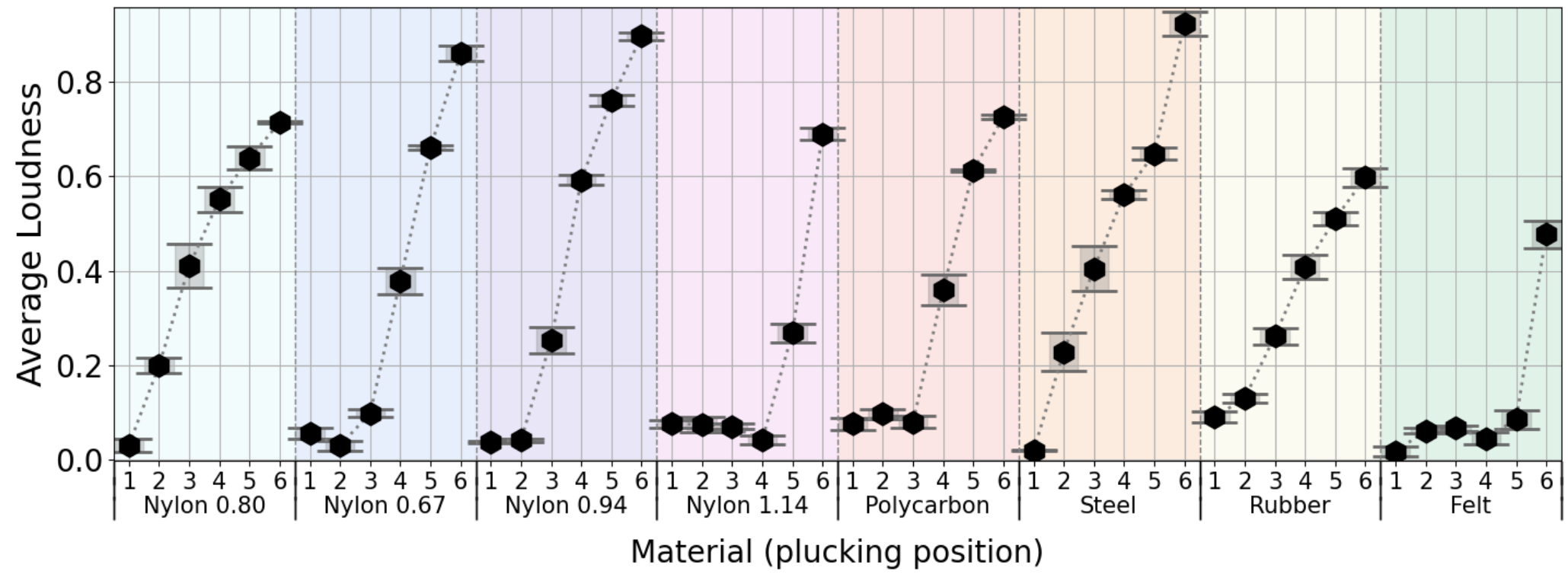

**Figure 6.** Comparison of average loudness values for measurement series conducted for each plucking position with all plectra.

It is clearly visible that an increase in plucking depth causes an increase in the average loudness of the generated pluck. This is also true when analyzing the maximum loudness of the sound as well as the RMS of the recorded signal.

These results also show a trend that will remain true for most other parameters. The experimental results for certain plectra (nylon 0.67, nylon 0.94, nylon 1.14, polycarbonate and felt) exhibit a shelf in the lower plucking depths. The changes in depth do not result in an extensive change in the produced sound, despite the string being clearly plucked at each depth. This is true until a threshold is passed after which an increase in depth results in a monotonic and consistent change in sound parameters. Analyzing the results also shows that after crossing the beginning plateau the steepness of the graph differs sizably between different materials.

Another set of parameters worth investigating is a set of parameters describing signal energy and noise. Fig. 7 shows how spectral entropy changes in the recorded signals. Spectral entropy is calculated using the following formula:

$$H_s = -\sum_{i=1}^{N} P(f_i)\log_2 P(f_i), \quad (1)$$

where $P(f_i)$ is the normalized power frequency spectrum at frequency $f_i$ .

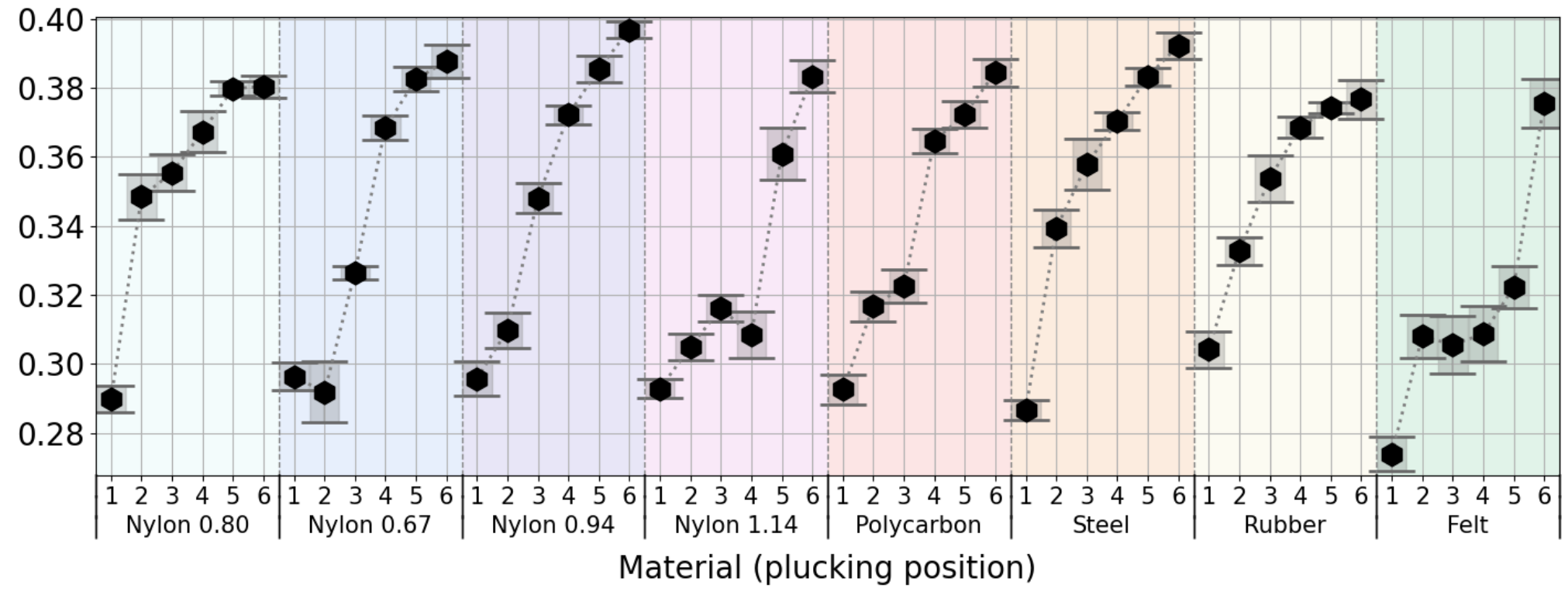

**Figure 7.** Comparison of signal spectral entropy values for measurement series conducted for each plucking position with all plectra.

This feature measures how evenly distributed the spectrum of a sound is. We can see that spectral entropy increases with a deeper plucking position signifying that the sounds spectrum becomes more evenly distributed with more harmonics being induced. It is also worth noting that if background noise were a large factor in the recorded signal, then the entropy for quiet plucks would be higher due to the high entropy of a noise spectrum.

To supplement these results Fig. 8 presents the values of zero crossing rate of the recorded signals. This parameter measures how often the signal changes sign, crossing zero. It is worth analyzing as it is a measure of the noisiness of the signal as well as it has been shown to be a strong correlate to listener perception of the sound's activity [19] defined as corresponding to the perceptual space described by the semantic pairs hard – soft, strong – weak, high energy – low energy and warm – cold as well as to the sounds brightness.

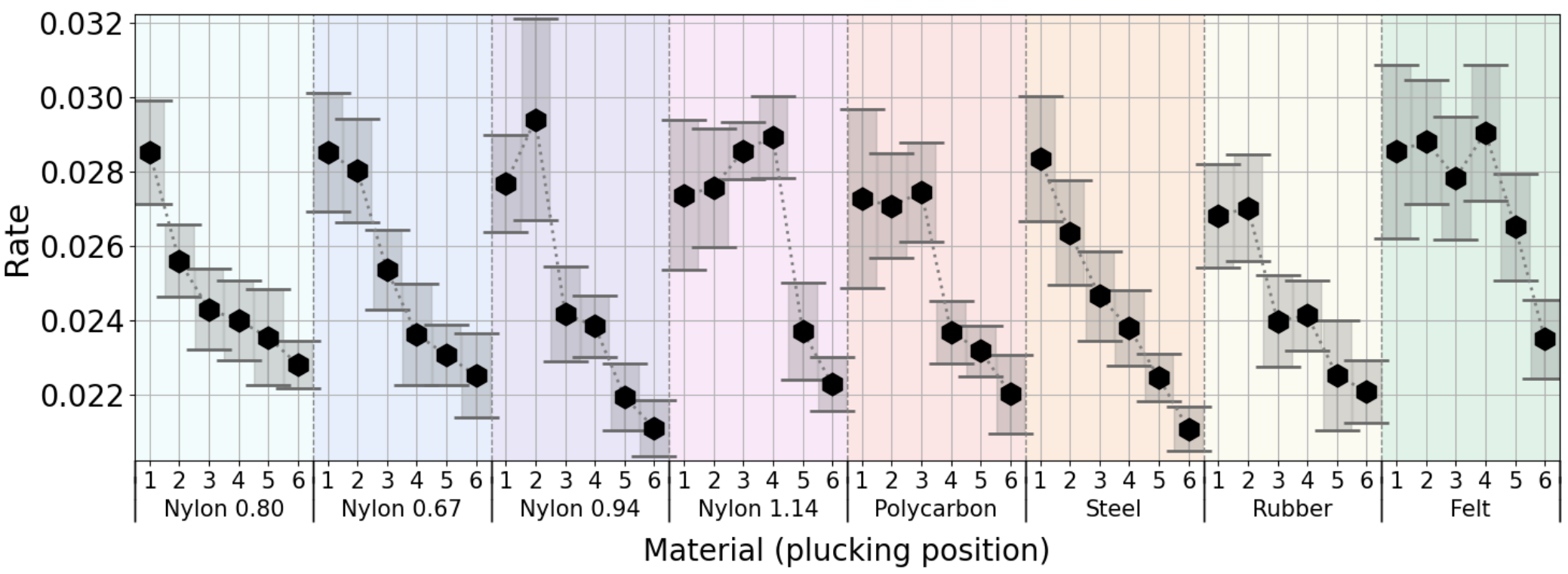

**Figure 8.** Comparison of zero crossing rate values for measurement series conducted for each plucking position with all plectra.

The results show that an increase in the depth of the plucking trajectory causes a drop in the zero crossing rate. This is consistent with previous results as an increase in harmonic content in proportion to noise as well as lower frequencies becoming dominant cause a less frequent crossing of zero.

The presented results show that as the plucking depth increases the sound becomes louder, while also becoming fuller and colder in timbre.

### 4.3. Spectral parameters

Further understanding can be gleamed through analysing a set of spectral features. The first one is spectral centroid, which is a measure of the centre of mass of the spectrum of the sound. It has been shown to be highly correlated to perceived brightness [17–20]. Fig 9 presents the values of spectrum centroid.

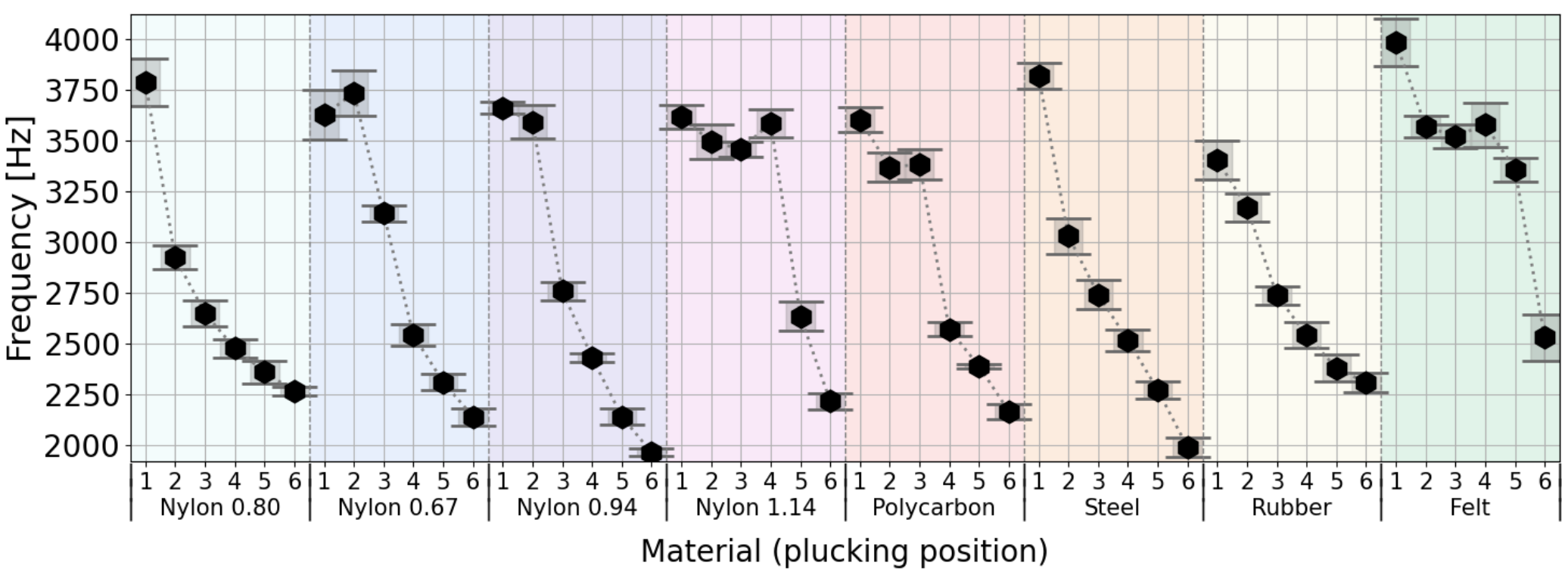

**Figure 9.** Comparison of spectral centroid values for measurement series conducted for each plucking position with all plectra.

Results show that increasing the plucking depth results in a fall of the spectral centroid after crossing a certain plucking depth threshold. The sound produced by the instrument becomes less bright. This is due to an increase of magnitude in low frequencies harmonics, which become dominant in the recorded signal. This is in line with the results presented in earlier sections. The described trend repeats for the following parameters: spectral spread, spectral bandwidth and spectral rolloff.

Additionally we can observe the rate of change of the spectra through the analysis of spectral flux, shown in Fig. 10.

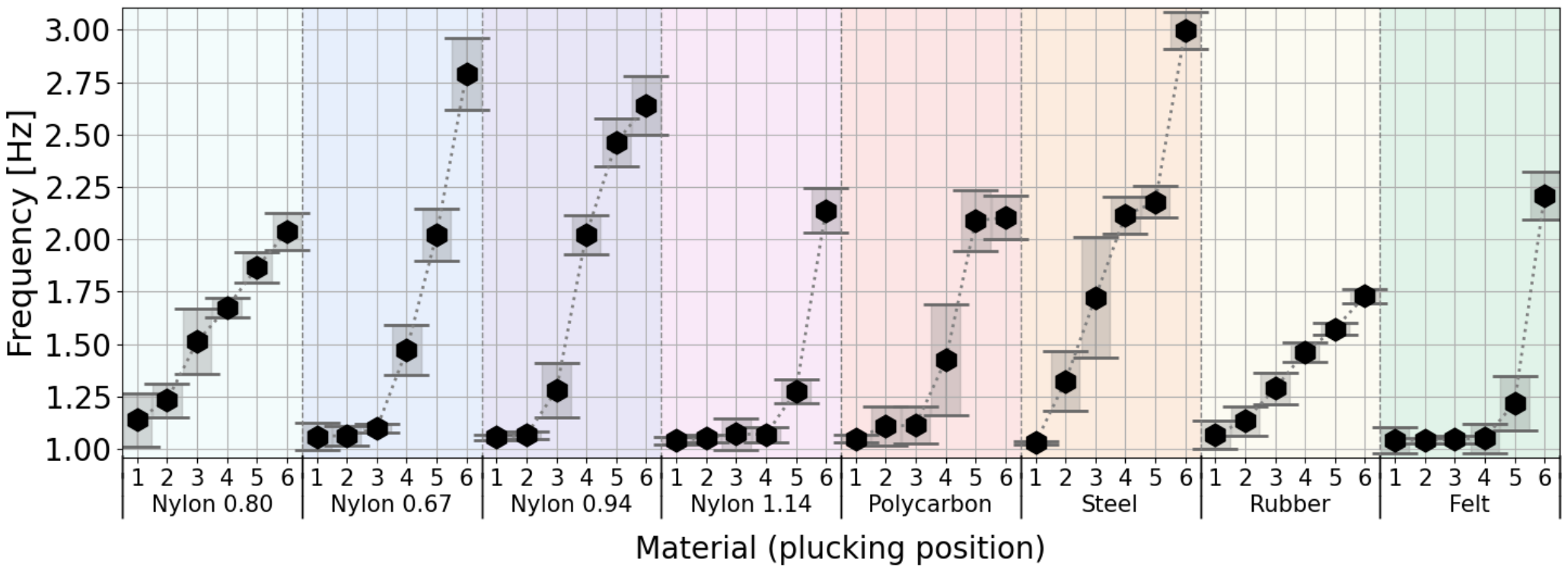

**Figure 10.** Comparison of spectral flux values for measurement series conducted for each plucking position with all plectra.

These results show that beneath the threshold depth for each material the flux is very low and remains stable. This changes however with a further increase of the plucking trajectory where the calculated flux rises sharply. Two things are worth pointing out. Firstly, in these results the difference in the rate of change is most visible. Materials such as rubber and nylon 0.80 have a much slower rate of rise than steel. Secondly, we can notice certain hitches in the evenness of the parameter change. In both steel and polycarbonate after crossing the lower threshold there is an even rise, which is disrupted, in polycarbonate between positions 5 and 6 and in steel between positions 4 and 5.

### 4.4. Harmonic feature analysis

Another aspect worth investigating are harmonic features which analyses the magnitude of harmonics and the proportions between them. Two features of particular interest are tristimulus 2 and tristimulus 3, which are defined by the following formulas:

$$Tristimulus\ 2 = \frac{a_2 + a_3 + a_4}{\sum_{h=1}^{H} a_h}, \tag{2}$$

$$Tristimulus\ 3\ =\ \frac{\sum_{h=5}^{H} a_h}{\sum_{h=1}^{H} a_h}, \tag{3}$$

where $a_h$ is the magnitude of the $h$-th harmonic, and $H$ is the total number of partials considered [20].

They have been shown to be highly correlated with the perceived roughness or harshness and sharpness of the sound with these features increasing with the rise of tristimulus 3 and decreasing with an increase of tristimulus 2 [21]. Figure 11 presents the values of both parameters.

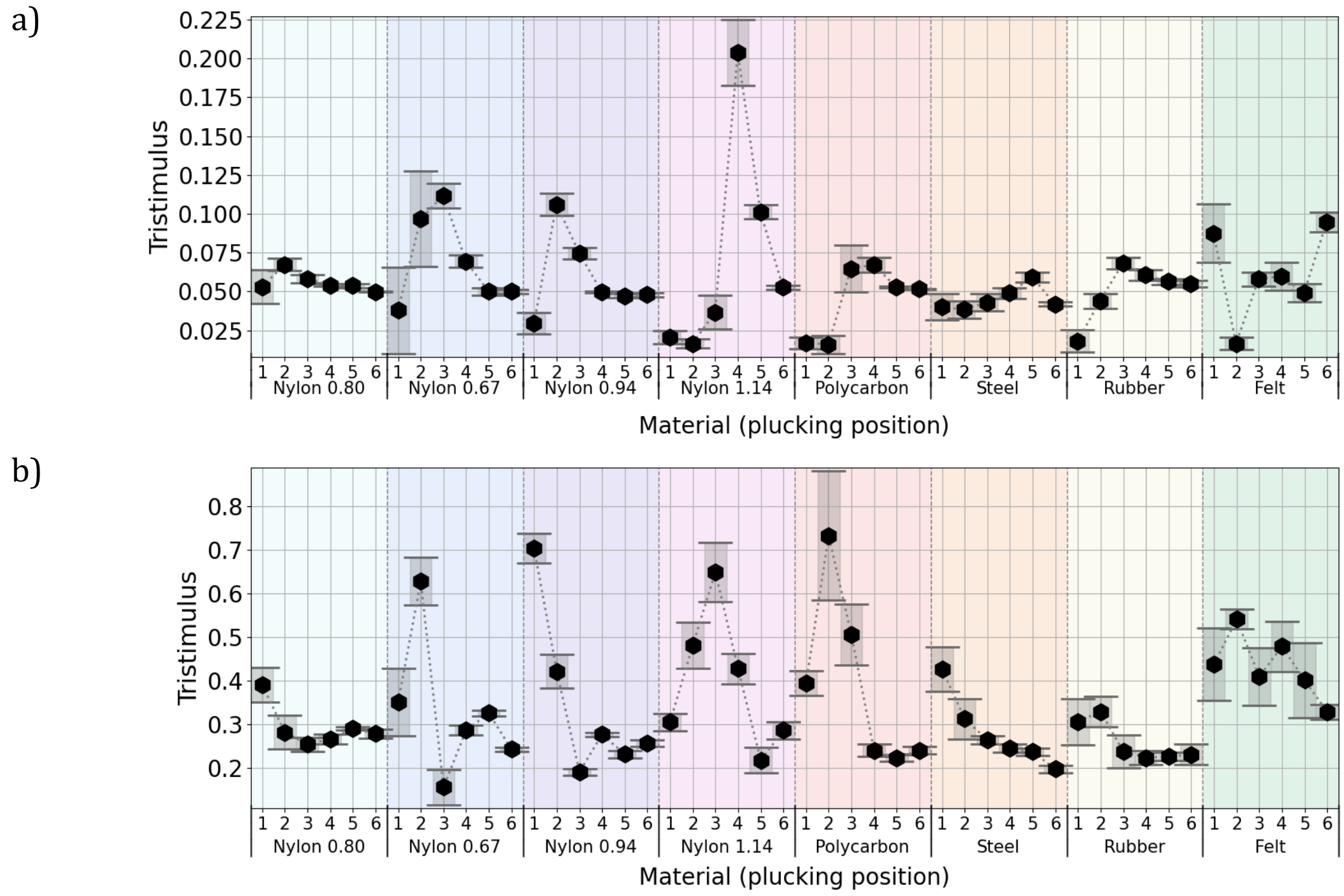

**Figure 11.** Comparison of a) tristimulus 2 and b) tristimulus 3 values for measurement series conducted for each plucking position with all plectra.

In contrast to previous parameters the calculated values do not follow a uniform monotonic change, with the trends being more complex. Due to the tristimulus 2 formula [20] it is sensitive to small changes in the magnitudes of the 2, 3, and 4 harmonics. This means that small changes can cause a large shift in its values, which can be seen for nylon 1.14 mm, nylon 0.94 and felt with sudden jumps between steps. It is worth noting that for most materials the values of tristimulus 2 rise during the first few positions and then drop towards the final ones, with only felt not following this trend. This shows a more complex relation between timbre and plucking trajectory than described earlier, where the values of features rose monotonously. This is even more true for tristimulus 3 where the general trend is towards a decrease along with an increase in plucking depth, but values can differ vastly between positions. While certain more general trends describing the instruments timbre seem to be straightforward the exact proportions between the magnitude of harmonics are subject to strong changes as they can be shifted by even minimal changes during the plucking process. This is important as the perception of sound is particularly sensitive to the relation between partials and the dissonant or consonant character that dominates in a sound. It is also worth noting that despite perceived harshness and sharpness being inversely correlated with respectively tristimulus 2 and 3, the presented results do not show them to be closely inversely correlated.

## 4.5. Timbre evolution during sound decay

While single number features serve as an overview of the entire sound it is also worth investigating how timbre changes along with the decay of the instruments sound. Figure 12 shows the change of spectral centroid over time for different plucking positions.

Firstly, as the guitar's sound decays the spectral centroid value rises. Decaying sound becomes quieter, therefore there might be two reasons for the rise of the centroid. One could be internal, related to different sound radiation characteristics for various frequency bands due to the instrument design and materials used. Another one could be the background noise which becomes a larger part of the sound as it decays. The noise might be reduced in future studies using an improved measurement system, which would allow to observe more details of the radiation characteristics. Secondly, it is worth noting that with an increase in plucking position the value curve becomes closer to each other. This supports the theory that below a certain threshold depth the string is not properly excited, and the sound does not decay as expected. Thirdly, for positions 3-6 one can see a sharp drop at the beginning as the sound changes from a spectrally wide impulse to a stable harmonic sound dominated by low frequencies. Fourthly, despite the graphs being smoothed using a moving window, it is still clearly visible that the decay is not monotonic with the spectral centroid increasing and decreasing. This shows how energy within the instruments can be transferred between different elements and their vibration, resulting in shifts of the finally produced sound.

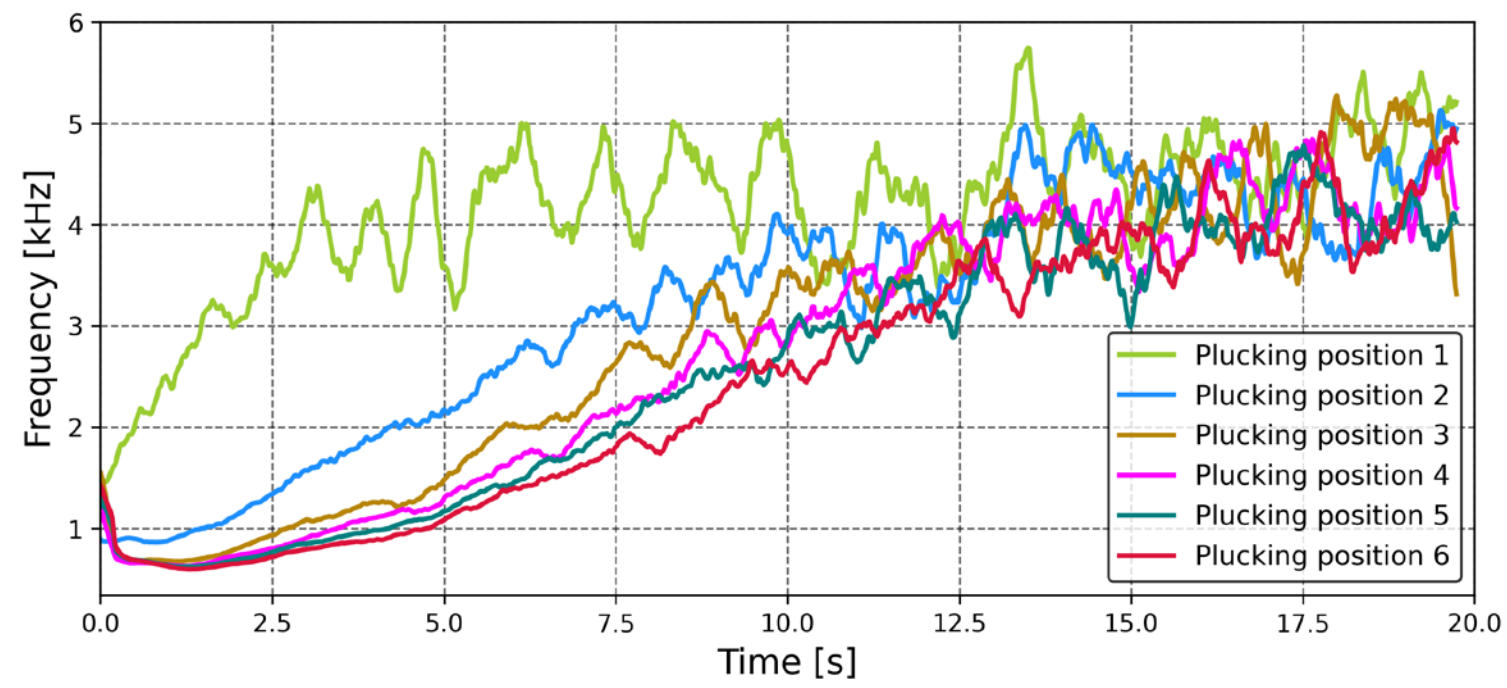

**Figure 12.** Comparison of spectral centroid values over time for each plucking position with the nylon plectrum.

### 4.6. Temporal analysis

To complete the look into the sound of the instrument we will present the effect plucking trajectory has on its decay time. Figure 13 presents a comparison of the RMS amplitude envelopes of decaying sounds plucked for all plucking positions.

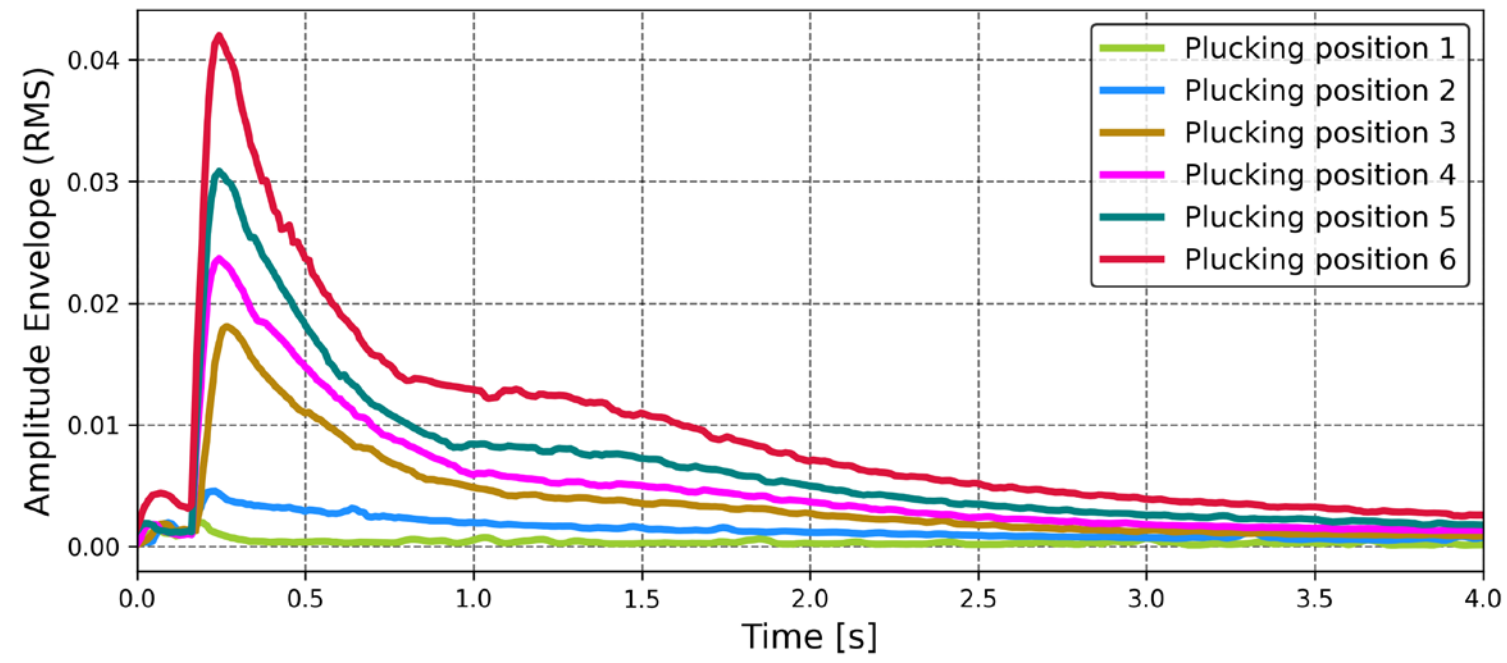

**Figure 13.** Comparison of RMS amplitude envelopes for each plucking position with the steel plectrum.

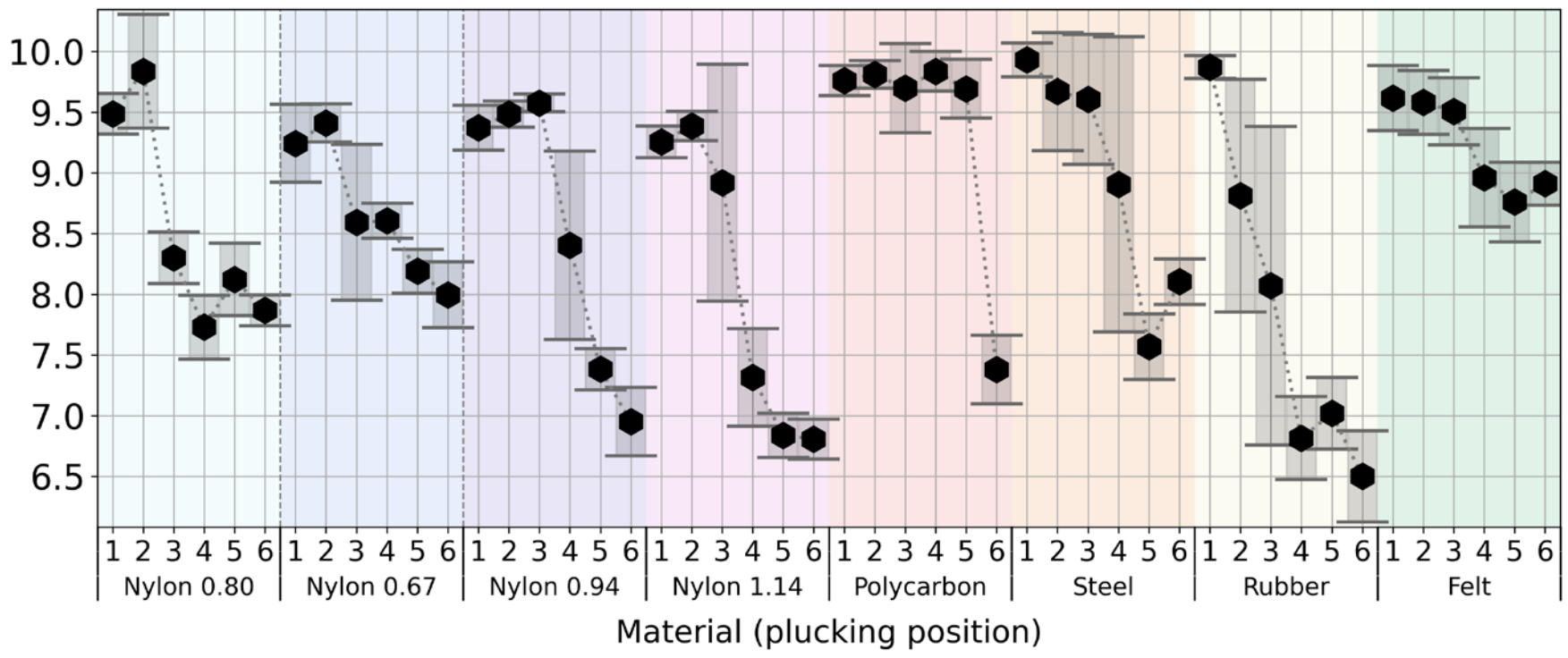

**Figure 14.** Comparison of temporal centroid values for measurement series conducted for each plucking position with all plectra.

As the plucking depth increases past the second position, the maximum magnitude of the envelope increases, while the shape of the decay curve remains very similar. After the sharp rise due to the pluck, we see a steep decrease for the first second, which becomes more gradual after that. Figure 14 shows the temporal centroid of the magnitude envelope, showing more generally how the time characteristic shifts.

As the plucking depth increases, the energy in the pluck also rises, bringing the values of the centroid down, towards the beginning of the recording. This is noticeable for all materials, though it is worth noting that for nylon 0.8 and steel there are points breaking away from this trend with a magnitude of change larger than the standard deviation showing that the proportions between peak loudness and decay time are not always exactly the same.

### 4.7. Experimental repeatability

A critical aspect of any experimental results is our confidence in the repeatability and reproducibility of the experiment as this sets the boundaries of what conclusions can be drawn. In this research we must distinguish two different aspects to this. Firstly, how repeatable the used plucking mechanism is in creating a series of plucks. Secondly, whether series conducted for different plectrum materials are comparable. While the experiment was designed to investigate the influence the plucking trajectory has on the instruments sound, it is also worth investigating whether conclusions regarding comparisons between different plucking materials can be confidently drawn. To investigate this an experimental series was conducted three times for a single plectrum with it being removed and mounted once again between each measurement. The results of this comparison are presented in Fig. 15.

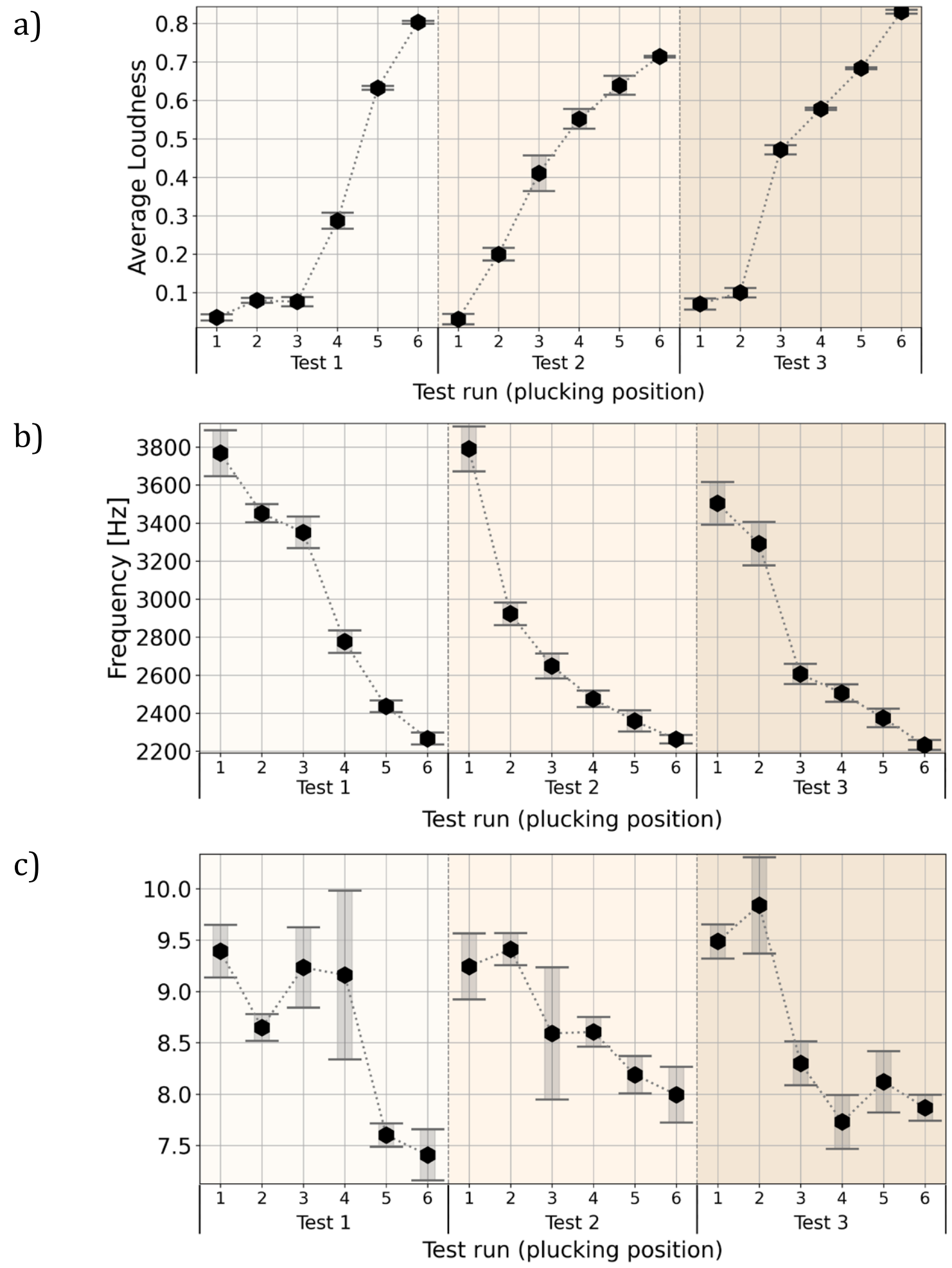

**Figure 15.** Comparison of a) average loudness, b) spectral centroid and c) temporal centroid values for measurement series conducted for each plucking position repeated for a nylon 0.8 mm plectrum.

First, we observe that the standard deviation of any given series of measurements is relatively low giving us confidence in the plucking mechanism. The comparison between test runs, however, shows sizable discrepancies. This is due to the mounting of the plectrum. Despite all efforts the process of unmounting and reaffixing the plectrum causes the results to not be repeatable. This shows how sensitive the guitar is to very small shifts in the position of the plectrum when plucking the string.

All this means that while we can be confident in the results obtained within a single measurement series conducted for a plectrum, the conducted experiment cannot be used to draw any conclusions about how the sounds created by different plectra compare. Such an experiment would require a very high level of precision together with a way of considering the differences in material properties, shapes and thicknesses of different picks.

## 5. Discussion

The results show that the increase in plucking depth causes the sound to become less bright and fuller with a decrease in perceived sharpness and roughness. While it could have been expected that increase in depth increases the magnitude of harmonics, it is interesting to see that this rise is proportionally larger for low frequencies. Moreover, if a pluck is too light, it fails to excite the lower harmonics of a string.

An observation that was not expected is a presence of a "shelf" in graphs of many of analysed parameters. The analysed value does not change with increasing plucking depth until reaching a certain point, where the change begins with a different steepness depending on the material. This behaviour can be observed for average loudness and spectral flux, but not in steel or rubber plectra, and one of nylon plectra. It is also visible, albeit weaker, in zero crossing rate and spectral centroid.

One set of parameters behaves differently: it neither shows a monotonic change, or a shelf. It is the tristimulus 2 and 3, which for the most cases displays an increase followed by an immediate decrease, over the area of six measured depths. Tristimulus (1, 2 and 3) was designed and might be interpreted as an analogue to a colour vision and primary colours. Tristimulus 2 and 3 behave differently for different materials, which might point to their usefulness due to sensitivity to this kind of change, which is not observed to the same degree in other parameters. One might observe the flow of maximum values between different tristimuli to determine the material of a plectrum.

Regarding the main subject of the study, it is safe to state that the majority of analysed sound properties is sensitive to changes in plucking trajectory as small as 192 μm. This is in line with a common knowledge among musicians that even a slight, delicate alteration in articulation produces perceptible results. As a consequence, while studying properties of plucked chordophones it is crucial to use highly repeatable mechanisms and procedures.

While for the most studied cases data deviation is low among 10 averaged plucks, the test of repeatability of mounting conditions proved that either the stand requires improvements, or an additional procedure is required to ascertain a common starting point. Without it one should not draw conclusions regarding absolute comparisons between separate mounts of studied plectra. It might be considered one of the most important findings of the study. It shows that a weak point is not the repeatability of subsequent excitations, which is often adequate, but the mounting of instrument and its various elements. They are not naturally adapted for fixed mounting, but also should not be modified for such mounting, because it could affect properties of the original sound. A more proper way to address this issue seems to be designing a procedure that would consider more advanced acoustic indicators of a common starting point.

## 6. Conclusions

The study presented in this manuscript explores the impact of micro-changes in the plucking trajectory of a guitar pick on the sound of an acoustic guitar. A test stand equipped with a Cartesian coordinate robot designed for various kinds of string excitation was used to perform a series of measurements where a string plucking trajectory was adjusted with a step of 192 micrometers, resulting in an increased attack depth. Eight different plectra have been studied, each one plucking with six different depths. The effect of these changes on loudness, timbre, harmonic content and sound decay has been analysed.

The results of the study show that at a low depth the string is not fully excited resulting in weak and markedly altered sound. The range of this effect changes with the mechanical properties of the plectrum material. After this range an increase in depth results in an increase in sound loudness, a decrease in inharmonicity and noisiness and a shift in timbre where the sound becomes fuller in low frequencies and rougher. Steel and rubber plectra behave differently than nylon, polycarbonate and felt. In their case parameters start to change immediately with increasing plucking depth, which is not the case with three remaining materials, where a kind of "shelf" can be observed in parameter graphs. Tristimulus parameters seem to be particularly useful in distinguishing plectra made of different materials.

Presented findings help to understand the nuanced relationship between plucking trajectory and acoustic output. They provide important insights regarding the importance of plucking in guitar testing methodologies, showing that the mechanics of plucking must be taken into account when conducting and interpreting results of guitar testing – even a small, 192 micrometer change in plucking depth alters signal parameters significantly. These findings agree with general conclusions of other studies regarding string excitation phenomena. All studies confirm that sound properties are highly sensitive to the character of string excitation in case of excitation mechanism parameters [3, 4], plucking position [8], plectrum thickness [10, 11], and plucking angle [12]. However, none of the previous studies regarded influence of micro-changes in plucking trajectory or depth on multiple sound characteristics. Therefore findings presented in this study are novel and were possible thanks to application of a Cartesian coordinate robot.

The test stand and the robot used for carrying out measurements performed satisfactory when a single series of measurements was considered. However it occurred that remounting a plectrum resulted in data shifts, so the initial plucking point was not reproduced well enough. There are two possible ways to solve the issue. One is to enhance the mounting of an instrument, its elements, and a plectrum. This however, might impact the original sound of a guitar by reducing instrument vibrations. Another way is to apply a procedure that would establish initial plucking position based on more advanced acoustic indicators.

## Additional information

The authors declare: no competing financial interests and that all material taken from other sources (including their own published works) is clearly cited and that appropriate permits are obtained.